\documentclass[12pt,a4paper,final]{iopart}

\usepackage{iopams}
\usepackage{graphicx}

\usepackage[breaklinks=true,colorlinks=true,linkcolor=blue,urlcolor=blue,citecolor=blue]{hyperref}

\begin{document}

\title[Conductivity fluctuations in proton-implanted ZnO microwires]{Conductivity fluctuations in proton-implanted ZnO microwires}

\author{B.~Dolgin$^{1}$}

%\ead{dolginn@post.bgu.ac.il}

\author{I.~Lorite$^2$}
%\ead{lorite@physik.uni-leipzig.de}
\author{Y.~Kumar$^2$}
%\ead{yogesh.kumar@physik.uni-leipzig.de}
\author{P.~Esquinazi$^2$}
\eads{\mailto{esquin@physik.uni-Leipzig.de}}
\author{G. Jung$^1$}
%\ead{jung@bgu.ac.il}
\author{B.~Straube$^{3,4}$}
%\ead{bstraube@herrera.unt.edu.ar}
\author{S.\,Perez~de~Heluani$^4$}
%\ead{SPerez@herrera.unt.edu.ar}

\address{$^1$Department of Physics, Ben Gurion University of the Negev, P.O.B. 653, 84105 Beer Sheva, Israel}
\address{$^2$Division of Superconductivity and Magnetism, Institute for experimental Physics II,
Fakult\"at f\"ur Physik und Geowissenschaften, Linn\'estra{\ss}e 5, 04103 Leipzig, Germany}
\address{$^3$CONICET, Dpto.~de F\'isica, Facultad de Ciencias Exactas y Tecnolog\'{i}a,
Universidad Nacional de Tucum\'{a}n, Argentina}
\address{$^4$Laboratorio de F\'isica del S\'olido,  Dpto.~de F\'isica, Facultad de Ciencias Exactas y Tecnolog\'{i}a,
Universidad Nacional de Tucum\'{a}n, Argentina}

\bibliographystyle{iopart-num}

\begin{abstract}
The electric noise can be an important limitation for applications
of conducting elements of size in the nanometer range. The
intrinsic electrical noise of prospective materials for
opto-spintronics applications like ZnO has not been characterized
yet. In this study we have investigated the conductivity
fluctuations in 10~nm thick current paths produced by proton
implantation of  ZnO microwires at room temperature. The voltage
noise under a constant dc current bias in undoped as well as in
Li-doped microwires is characterized by $1/f^a$ power spectra with
$a \sim 1$. The noise intensity scales with the square of the bias
current pointing out to bias-independent resistivity fluctuations
as a source of the observed noise. The normalized power spectral
density appears inversely proportional to the number of carriers
in the probed sample volume, in agreement with the
phenomenological Hooge law. For the proton-implanted ZnO microwire
and at 1~Hz we obtain a normalized power spectral density as low
as $\sim 10^{-11}~$Hz$^{-1}$.
\end{abstract}

%\vspace{2pc} \noindent{\it Keywords}: noise, semiconductors,
% defects
\submitto{\NT}
%\maketitle

\section{Introduction}

The quest for further reduction of the components' size of
electronic devices can be seriously limited by an increase of the
$1/f$ electric noise, particularly in  metallic nanowires
\cite{bid06}. However, the phenomenon of defect-induced magnetism
(DIM), i.e., magnetic order produced by defects and/or added ions
(not necessarily magnetic) in the atomic lattice of oxides and
other materials, indirectly requires a pronounced sample size
reduction
\cite{Peng:PRL01,McCluskey:JAP09,Xing:AIP11,Esquinazi:I13,Guglieri:ADFM14,A1,A5}.
Due to the O-2$p$ hole-spin polarization from oxygen atoms around
Zn-vacancies in ZnO\cite{Yin:PRL10,lor15apl}, the DIM phenomenon
opens up new possibilities for spintronics applications of the ZnO
compound. Due to the very nature of DIM and the difficulties in
creating an adequate density of defects that would be
homogeneously distributed all over the entire volume of an oxide
sample,  different experimental strategies were used to obtain
large ferromagnetic magnetization values at saturation. These
include, for example, ion irradiation \cite{Esquinazi:I13,zho14},
 co-doping, \cite{her10,yi10,lor15apl} and sample size
reduction \cite{lor15apl,lorjp15}.  By implantation of protons at
energy $\lesssim 300~$eV one can obtain a magnetically ordered
$\lesssim 10~$nm~thick surface regions in ZnO:Li microwires. These
regions show clear ferromagnetic characteristics, as revealed by
SQUID magnetometry and x-ray magnetic circular dichroism
(XMCD)\cite{lor15apl} as well as a negative magnetoresistance
\cite{lorjp15}. Therefore, increase of the noise associated with
decrease of the effective size of the magnetically ordered sample
volume may become a limiting factor for the applications of
magnetic oxide nanowires in spintronics.

Let us underline that the electric noise measurements reported so
far for ZnO nanowires were performed using nanowires incorporated
into field effect transistor structures\cite{xio08,ju08}. For this
reason and in most cases, such measurements   do not necessarily
reveal the intrinsic noise of the nanowires but rather the
dominating interfaces and contacts contributions. The main aim of
the experimental work we report in this letter is the
investigation of the intrinsic conductivity noise spectra of ZnO
microwires implanted with protons, in which basically only a very
thin, nanorange sized surface layer of the wire conducts the
electricity.

We have investigated the transport noise properties of two ZnO
microwires at room temperature. The first ZnO wire is doped with
Li and shows magnetic order at room temperatures after proton
implantation \cite{lor15apl}. For comparison, a second ZnO wire
without Li doping but with a similar proton implantation dose as
the first one, has also been measured. This latter wire does not
show any magnetic order at room temperature\cite{lorjp15}. Both
microwires are characterized by $1/f$-like noise. Its intensity is
several orders of magnitude smaller than that reported for ZnO
nanowires in field effect transistors\cite{xio08,ju08} and purely
Gaussian statistical properties.

\section{Experimental details}

Clean and Li-doped ($\simeq 3\%$) ZnO powders were obtained by
thermal decomposition of Zinc Acetate. As starting precursors Zinc
Acetate dihydrate Zn(CH$_3$COO)$_{2}\cdot$ 2H$_{2}$O and
Li-hydroxide mono hydrate LiOH$ \cdot$ H$_{2}$O of (Sigma-Aldrich)
(99.99\%) commercial chemicals were used. The precursors were
mixed in 30~ml of tri-distilled water at a nominal concentration
of 3~at.\%~ Li/Zn.   The obtained wires, i.e. pure ZnO and
Li-doped ZnO showed wurzite like structure with hexagonal
morphology with the c-axis as main wire growth axis. Further
details of the microwire preparation can be taken from
Refs.~\cite{lor15apl,Villa:JAP14}.

\begin{figure}
\centering
 \includegraphics[width=1.0 \linewidth]{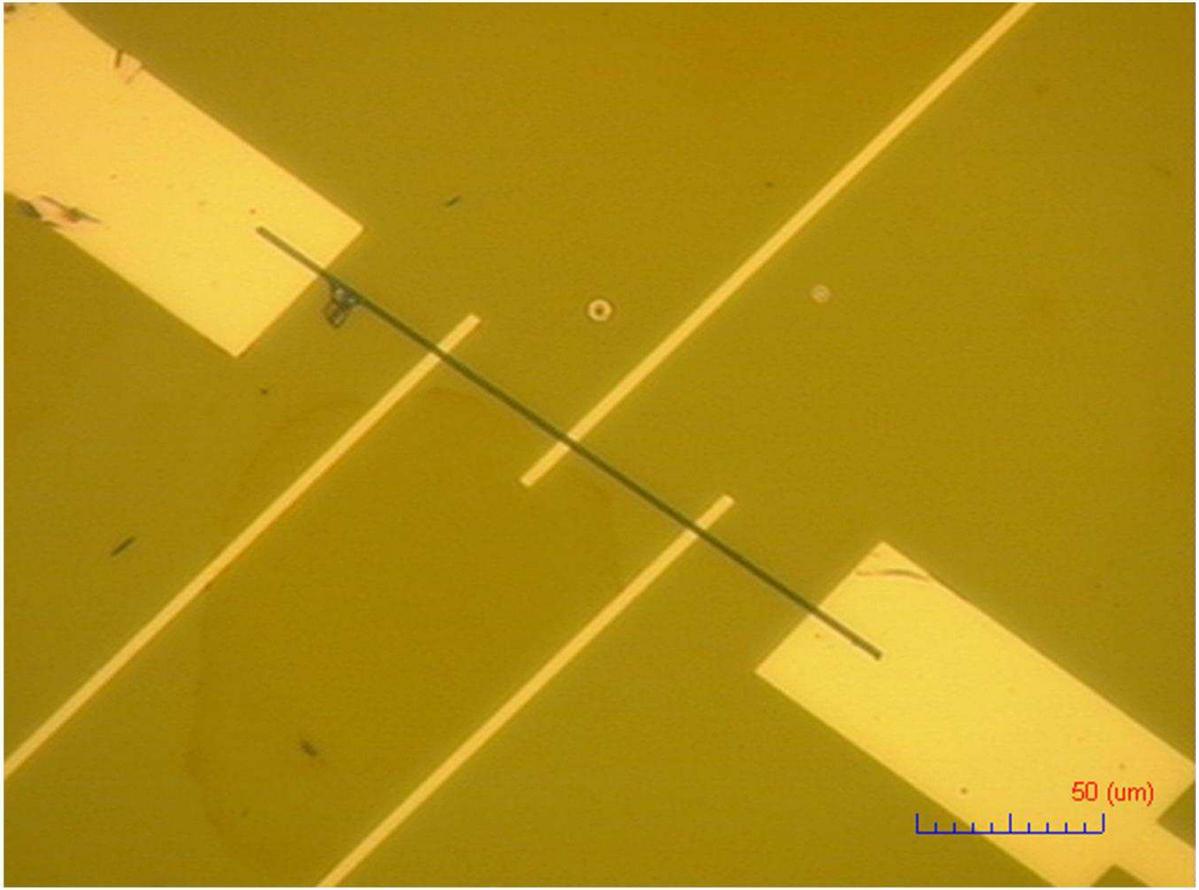}
  \caption{Optical picture of the Li-doped ZnO nanowire
  after hydrogen implantation (H:LiZnO) with the five in-line Au contact electrodes.}
  \label{con}
\end{figure}

The size of the two wires were: (diameter $\times$ length between
the voltage measuring electrodes) $15 \times 400~\mu$m$^2$ for the
ZnO microwire (ZnO) and $0.8 \times 80~\mu$m$^2$ for the Li-doped
ZnO nanowire (LiZnO), see Fig.~\ref{con}. The pure and the
Li-doped wires were exposed to remote H$^+$ DC plasma ($\lesssim
300$~eV H$^+$ implantation energy) in the parallel-plate
configuration (60~$\mu$A bias current for one hour at a chamber
pressure of 0.5~mbar) using a procedure described in
Refs.~\cite{kha11,Lorite:JMR13}. The ZnO wire was electrically
contacted with indium electrodes after the entire wire was proton
irradiated. The indium contacts were necessary because the large
diameter of the wire hindered simple lithographic contacts. No
annealing was necessary and the ohmic behavior was checked before
we started the noise measurements. In the case of the Li:ZnO
nanowire the proton implantation and the contacts were performed
in two steps. Firstly, the contact paths were done with electron
lithography and the proton implantation of the contact regions was
carried out, previous to any metal deposition. Afterwards, the
Pd/Au electrical contacts were deposited by sputtering on the
electrode paths. Following this procedure the electrical contacts
always showed good ohmic behavior. Finally, the proton
implantation on the rest of the wire was performed. This procedure
was necessary to avoid any diffusion of H$^+$ during the
lithography process.

 The resistivity of the wires
after H$^+$ implantation at 300~K were $\rho \simeq
6.8~\mu\Omega$cm ($0.22~\mu\Omega$cm) for H:LiZnO (H:ZnO), several
orders of magnitude smaller than before implantation. All Li-doped
ZnO microwires are highly insulating with a resistance larger than
$10^{10}~\Omega$ at room temperature \cite{lorjp15}. At the energy
used for implantation of H$^+$ in our samples, a significant
concentration of protons as well as defects (Zn- and O-vacancies)
are localized only within a $\sim ~ 10~$nm surface region
according to SRIM simulations \cite{Ziegler2} (see also Fig.~1 in
\cite{kha11}). The defects concentration (Zn-vacancies) is
important for the magnetism \cite{lor15apl,lorjp15}. In fact, it
was experimentally verified that most of the magnetic signal comes
from the near surface region \cite{kha11}.  The temperature
dependence of the resistance can be understood using a simple
parallel resistor model that takes into account two contributions
to the total conduction, one from the near surface region and the
other from a non-magnetic region further inside the wire, where a
lower H$^+$ doping also reduces, but in a minor grade, the
resistance \cite{lorjp15}. Transport studies of H$^+$-implanted
ZnO single crystals \cite{kha12} and microwires \cite{lor14}
indicate that the metallic-like contribution is associated with
the hydrogen-rich near surface region of the samples. The
photoluminescence spectra at 300~K and the magnetic properties
(SQUID, XMCD and magnetoresistance) of the wires have been
reported recently \cite{lor15apl,lorjp15}.

The conductivity noise was measured at room temperature by biasing
the samples with dc current, supplied by high output impedance
current source, and recording the resulting voltage fluctuations.
Due to the very high resistances of the wires we have employed a
five-point contacts arrangement to reduce the dc voltage drop on
the current biased sample, see Fig.~\ref{con}. In this contact
arrangement the current flows from the two most external contacts
to the grounded collector contact located at the middle of the
sample, between the two internal contacts across which the
fluctuating voltage is measured, see Fig.~\ref{con}. The voltage
drop across  the dc current biased wire was amplified by a
homemade very low noise preamplifier and processed by a
computer-assisted digital signal analyzer. The 5-contacts
arrangement allowed us to avoid problems associated with
preamplifier input saturation or with exceeding the common mode
rejection capability for all used input dc currents within the
range $10^{-3}~\mu$A~$ \le I \le 50~\mu$A. To eliminate
environmental interferences and noise background, along with each
measurement the power spectral density (PSD) at zero current was
measured separately and subtracted afterwards from the PSD
obtained at a given current flow, providing us the pure PSD of the
intrinsic sample fluctuations.

\section{Results and discussion}

\begin{figure}
%\begin{centering}
 \includegraphics[width=1.0 \linewidth]{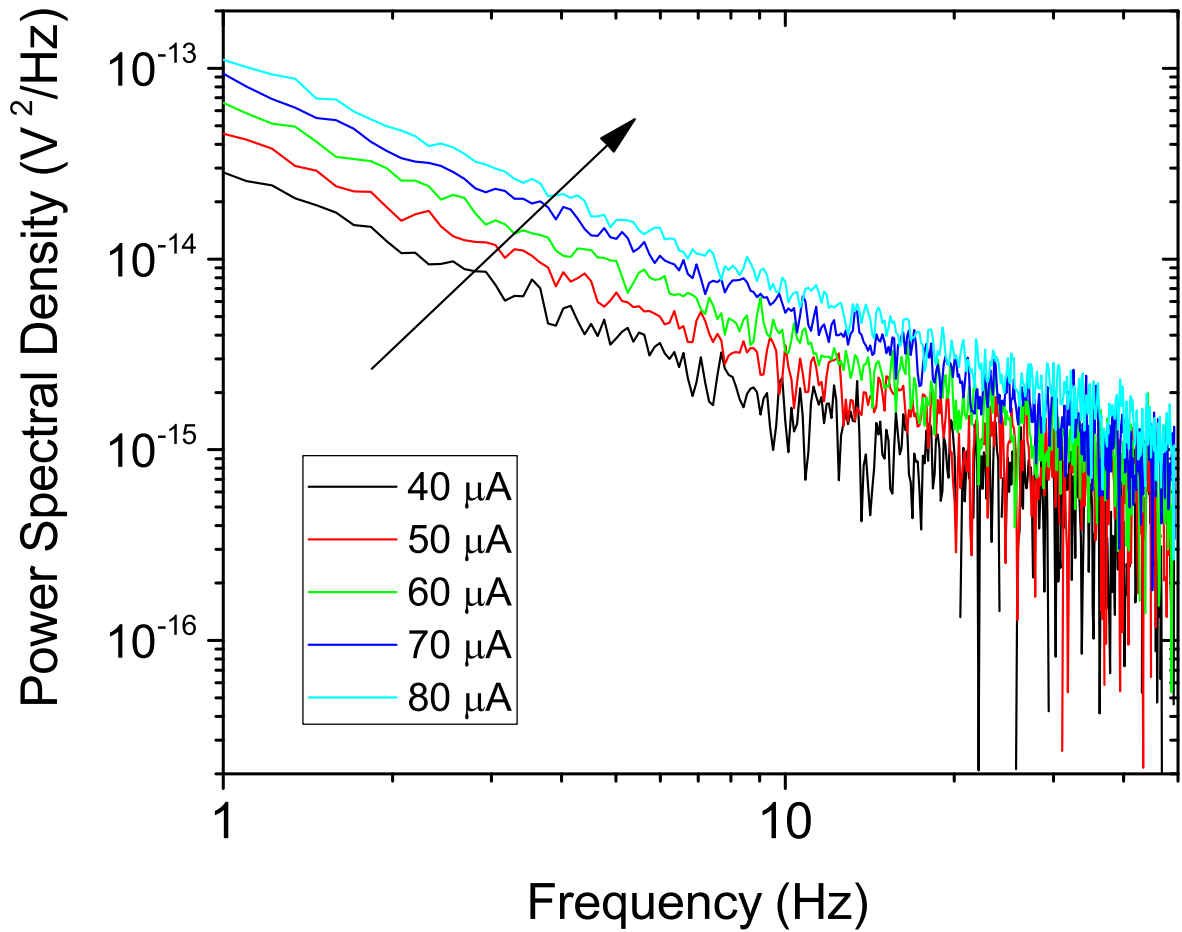}
  \caption{Power spectral density  of voltage fluctuations
  at different dc current flows across H:ZnO nanowire. The arrow indicates the growth direction of dc bias current.}
  \label{fig:1}
  %\end{centering}
\end{figure}

\begin{figure}
%\begin{centering}
 \includegraphics[width=1.0 \linewidth]{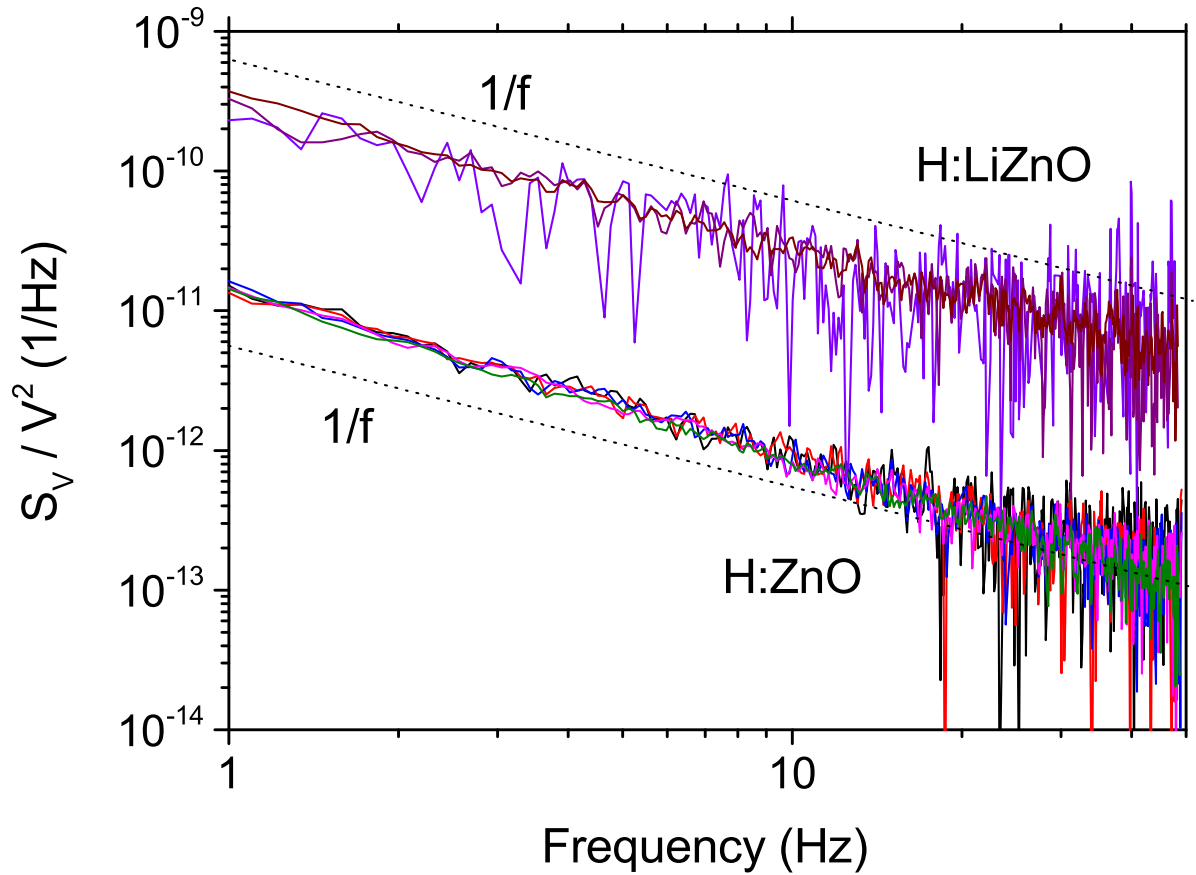}
  \caption{Normalized power spectral density (PSD) of
  voltage fluctuations for current biased H:ZnO and H:ZnLiO wires.
The different colors of the experimental curves mean curves taken
at different currents, see e.g. Fig.~\ref{fig:1}. The dashed lines
represent $1/f^a$ with the spectral exponent $a=1$. }
  \label{fig:2}
  %\end{centering}
\end{figure}

Figure \ref{fig:1} shows a set of averaged PSD recorded at
different current dc current flow in the H:ZnO microwire. The
spectra demonstrate a frequency dependence of the $1/f^a$ type
with the exponent $a \thickapprox 1$. The intensity of the noise
increases with increasing bias current.  Voltage noise with a
$1/f$ spectrum is generally related to resistance fluctuations,
which are measured by applying dc current and recorded as voltage
fluctuations. When the resistance fluctuations are just probed by
current, and are not influenced by its flow, then PSD of the
voltage noise scales as the square of the bias current. Such
modulation noise is frequently referred to as quasi-equilibrium
$1/f$ noise.  Figure~\ref{fig:2} shows PSD recorded under
different current flows and normalized by the square of the dc
voltage across the wire. For both samples, the normalized PSD
collapses to a single line indicating that PSD scales with the
square of the bias, meaning that the current flow does not excite
or influence these auscultations but only reveals them  by
converting conductivity fluctuations into the measurable voltage
noise. Consequently, $S_V/V^2=S_R/R^2$, where $S_V$ is the
measured PSD of voltage fluctuations and $S_R$ is the  PSD of the
underlying conductivity noise.

The measured noise does not depend  on weak, of the order of few
hundredths of Gauss, applied magnetic field and on illumination of
the wires with UV light. This is not  surprising for the undoped
H:ZnO wire because it is not magnetically ordered at room
temperature \cite{lorjp15}. For the magnetic one \cite{lor15apl}
the hysteresis loops with a finite width indicate either finite
pinning of domain walls and/or some unknown magnetic anisotropy.
With an applied field of the order of 0.5~kOe it is possible to
change the magnetization notably, probably due to the reduction of
the number of domains. The domain walls move back and forth at
room temperature, especially if there are AC magnetic fields or
currents applied to the wire. Any driven force would change
locally the magnetization and this might constitute a source of
additional Barkhausen-like noise. The frequency spectrum of this
noise can be very broad, up to the MHz region. Our study and the
negligible effect of the applied magnetic field on the measured
noise indicates that its source is not related to the magnetic
order of the sample.

The increase in the number  of carriers under UV light
illumination at room temperatures is small because both wires are
implanted with protons resulting in the resistance of the
implanted sections being much smaller that the resistance of the
untreated parts of the wire, which is most sensitive to the UV
light illumination. Because of this, most of the bias current is
shunted to the lower resistance part, which is only weakly
influenced by the light, with a relative decrease of the total
resistance under UV of the order of 10\% to 30\%. The UV  effect
on the electrical noise is therefore negligible. The absence of
any increase of the noise under UV is relevant for the detection
of defect-induced magnetic order in low-dimensional ZnO structures
\cite{lorsmall}.

There is, however, a noticeable  difference in the noise intensity
between the H:LiZnO and H:ZnO wires. The normalized noise, which
is insensitive to the actual value of the sample resistance, is 25
times higher in the H:LiZnO sample than in the H:ZnO. Moreover,
the PSD of the H:LiZnO wire is very close to the ideal $1/f$
noise. The spectral exponent obtained by fitting the experimental
spectra to $1/f^a$ form is $a= 1.08 \pm 0.015$, while the spectral
exponent of the H:ZnO wire noise is higher, $a= 1.30 \pm 0.02$.
Observe that the difference in the exponent $a$ is well beyond the
experimental uncertainty of the measurements.

To understand the difference in  the spectral properties of both
samples we have to recall the widely accepted nonexponential
kinetics model for $1/f$ noise in solids\cite{weissman}. According
to this model, $1/f$ noise results from incoherent superposition
of many elementary random two-level fluctuators (TLF), each
producing a Lorentzian contribution  with  a single relaxation
time $\tau$. The overall spectrum of a system is due to incoherent
superposition of contributions of all individual TLFs, and depends
on the distribution of the relaxation times $D(\tau)$ in the
system. When the distribution $D(\tau)\propto 1/\tau$, then the
integrated spectrum $S_V\propto \int{(\tau
D(\tau)/(1+\omega^2\tau^2)d\tau}$ becomes pure $1/f$ noise
$S_V\propto 1/\omega$. Assuming that the TLFs are thermally
activated, what in our case of the room temperature experiment is
quite obvious, we write for $\tau=\tau_0 \exp(E/kT)$, where
$\tau_0$ is the attempt frequency, usually related to the phonon
frequency in solids and $E$ is the activation energy in a
symmetric TLF. The resulting total noise spectrum of an ensemble
of active TLFs, given by
\begin{equation}
S(\omega)=
\int{\frac{\tau_0\exp[E/k_BT]}{1+\omega^2\tau^2\exp[2E/k_BT]}
D(E)dE},
\end{equation}
has a pure $1/f$ form for $D(E)=const.$  Therefore, the slope of
the PSD function,  i.e., the spectral exponent $a$, reflects the
shape of the energy distribution $D(E)$ within the experimentally
accessible energy window. A departure of the spectral slope from
$a=1$ implies a deviation of the energy distribution from a flat
one. In particular,  a spectral slope with $a>1$ implies an excess
in the density of the low energy fluctuators in the system.
Therefore, the distribution of energy barriers of the scatterers
in the H:LiZnO wire where $a$ is very close to unity is flat,
while that of the H:ZnO wire with $a=1.3$ is skewed in such a way
that there is an excess of scatterers at low energies. The reason
for such a difference can be intuitively understood remembering
that the introduction of protons to ZnO suppresses severely the
formation of compensating interstitials and enhances the acceptor
solubility in ZnO by forming H-acceptor complexes and leading to
enhanced conductivity\cite{Cheol:PRB04}. On the other hand, doping
with Li stabilizes and bounds hydrogen in the H:LiZnO structure
while in H:ZnO crystal the hydrogen is not bonded and diffuses
freely towards the surface and the interior. Therefore, the
distribution of hydrogen in the H:LiZnO is uniform resulting in a
flat distribution of activation energies while in H:ZnO the areas
close to the surface or at the interface with the insulating part
are enriched with hydrogen and have lower activation energies than
the remaining H-implanted parts of the wire resulting in a skewed
distribution of activation energies and consequently the spectral
exponent $a>1$.

The difference in the normalized noise magnitude  between the
wires can be ascribed to difference in number of charge carriers
$N$ in the effective volumes from which the noise has been
measured. The $1/f$ conductivity fluctuations in many solid state
systems can be quantified by a phenomenological Hooge equation
$S_V/V^2=b/Nf^a$ ($b$ is a constant) indicating that the
normalized noise at 1~Hz should scale inversely proportional to
the number of carriers $N$ \cite{weissman}. One can evaluate the
ratio of the number of carriers in the wires as ${N_{2}}/{N_{1}}$
(subindex 1 for H:LiZnO and 2 for H:ZnO)  from the ratio of the
measured resistances of both wires $R_{2}/R_{1}$, using the simple
Drude model in which the resistivity $\rho=m^*/ne^2\tau$, where
$m^*$ is an effective mass of charge carriers, $e$ their charge,
$\tau$ the mean time a charge carrier has traveled since the last
collision (the relaxation time), $n=N/V$ carrier concentration and
$V$ is the probed volume of the sample. The resistance of the
sample reads $R=\rho L/A=m^*L^2/Ne^2\tau$, where $L$ is the
distance between the contacts and $A$ the cross-section area of
the wire. From the last expression and considering similar
effective masses for both wires we have
\begin{equation}
\frac{N_{2}}{N_{1}}=\frac{R_{1}}{R_{2}}\frac{L^2_{2}}{L^2_{1}}\frac{\tau_{1}}{\tau_{2}}
\end{equation}
With the experimental values $R_{2}/R_{1}$=11.3 and
${L^2_{2}}/{L^2_{1}}$=25 one obtains
\begin{equation}
\frac{N_{2}}{N_{1}}=283\frac{\tau_{1}}{\tau_{2}}.
\end{equation}
Assuming that the noise intensity  ratio $S_{1}/S_{2}=25$ properly
reflects the ratio of number of carriers in the active volumes of
the two samples, to keep the consistency between the estimations
one has to assume that ${\tau_{1}}/{\tau_{2}}\sim 0.1$. This
appears to be a reasonable assumption taking into account  the
measured resistivities, the expected similar carrier densities in
the two wires (due to similar proton doses), and the Matthiesen
rule where $\tau^{-1} = \tau_{\rm phonons}^{-1} + \tau_{\rm
impurities + defects}^{-1}$. Due to the Li doping, which increases
the pinning of protons and Zn vacancies near Li ions, we expect a
clear decrease of $\tau_{\rm impurities + defects}$ for the Li
doped wire respect to the undoped one.

In conclusion, we have measured the intrinsic noise spectra of
proton implanted ZnO and Li:ZnO wires with a 10~nm thick
conducting surface layer.  The spectral density follows basically
a $1/f^a$ dependence with an exponent $a$ close to 1 for the
H:LiZnO wire. For the wire H:ZnO, however, the exponent $a \simeq
1.3$, which can be attributed to a non-constant distribution of
activation energies of the fluctuators assembly.  The consistency
between the resistivity ratio of the doped and undoped wires and
the ratio of the normalized noise intensity allows one to conclude
that the relaxation time in Li-doped ZnO microwires is an order of
magnitude shorter with respect to undoped samples. The intensity
of the normalized intrinsic microwire noise observed in our
experiments $\sim 10^{-11}~$Hz$^{-1}$,  orders of magnitude
smaller than the reported for ZnO nanowires in  field effect
transistor configurations\cite{xio08,ju08}. Therefore, it is clear
that in the latter configuration the observed noise is dominated
by extrinsic contacts and interface contributions.

Acknowledgments: This work was funded  by the Collaborative
Research Center SFB~762 ``Functionality of Oxide Interfaces'' in
Germany.

\bibliographystyle{apsrev}

\section*{References}
%\bibliography{lib,Halle_14,magnetic_carbon}

\newcommand{\noopsort}[1]{} \newcommand{\printfirst}[2]{#1}
  \newcommand{\singleletter}[1]{#1} \newcommand{\switchargs}[2]{#2#1}
\providecommand{\newblock}{}

\end{document}